\documentclass[12pt]{iopart}

\newcommand{\be}{\begin{equation}}
\newcommand{\ee}{\end{equation}}
\newcommand{\bea}{\begin{eqnarray}}
\newcommand{\eea}{\end{eqnarray}}
\newcommand{\jes}{j_{\hbox{\tiny ES}}}
\newcommand{\dx}{\hbox{d$x$}}

\begin{document}
\jl{01}

\letter{Coarsening in surface growth models without slope 
selection\footnote[8]{Dedicated to the Peanuts on the occasion of
their 50th birthday.}}

\author{Paolo Politi\dag\ddag\S\footnote[7]{Corresponding author.
Fax: +39 055 229330, E-mail: {\tt politi@fi.infn.it}} and 
Alessandro Torcini\dag\footnote[6]{E-mail: {\tt torcini@fi.infn.it},
URL: {\tt http://torcini.de.unifi.it/$\sim$torcini}} }

\address{\dag\ Istituto Nazionale per la  Fisica della Materia, 
Unit\`a di Firenze, L.go E.~Fermi 2, 50125 Florence, Italy}
\address{\ddag\ Dipartimento di Fisica, Universit\`a degli Studi di
Firenze, L.go E.~Fermi 2, 50125 Florence, Italy}
\address{\S\ Fachbereich Physik, Universit\"at GH Essen, 45117 Essen,
Germany}

\begin{abstract}
We study conserved models of crystal growth in one dimension 
[$\partial_t z(x,t) =-\partial_x j(x,t)$] which are linearly unstable 
and develop a mound structure whose typical size $L$
increases in time ($L \sim t^n$).
If the local slope ($m =\partial_x z$) increases indefinitely,
$n$ depends on the exponent $\gamma$ characterizing the large $m$
behaviour of the surface current $j$ ($j\sim 1/|m|^\gamma$): 
$n=1/4$ for $1\le\gamma\le 3$ and $n=(1+\gamma)/(1+5\gamma)$ for $\gamma>3$.
\end{abstract}


The conserved dynamics of a solid surface growing under the action of an 
external flux of particles is described by the continuum equation
\be
\partial_t z(x,t) = -\partial_x j(x,t) + \delta F(x,t)~,
\label{eq-z}
\ee
where $z(x,t)$ is the local height of the surface in a comoving frame
(so that the average value $\bar z$ is set to 0) 
and $\delta F(x,t)$ is the shot noise.

Thermodynamic and kinetic mechanisms contribute to $j$ and its actual
expression depends on the details of the growth process.
Here we are interested in the growth of a high-symmetry surface by
Molecular Beam Epitaxy (MBE),
where the instability has a purely kinetic origin: the reduced interlayer
diffusion~\cite{eh}. Nonetheless, our treatment will be as general 
as possible.

A wide class of models is described by the current
\be
j = K m''(x) + \jes (m)~,
\ee
where $m=\partial_x z$ is the local slope. 
The first term 
generally describes a thermally activated relaxation of the surface, 
but kinetic mechanisms can also contribute to $K$~\cite{KC}.

The second term is responsible for the instability and its origin is
an asymmetry in the sticking process of an adatom to a step
(Ehrlich-Schwoebel (ES) effect):
sticking from the upper terrace is hindered and this implies an up-hill 
current~\cite{VilJP} which is called Ehrlich-Schwoebel current ($\jes$).
Also other (generally stabilizing) processes can contribute to $\jes$
and this explains the different expressions $\jes$ may take~\cite{review}.

Whatever these processes are, $\jes$ is linear in $m$
at small slopes ($\jes\sim\nu m$) and therefore in the early stages of 
the growth it prevails on the first term ($Km''$) 
at sufficiently large wavelengths.  This means that 
the linear stability of the flat surface will be decided by the
sign of $\nu$, a positive one meaning instability. In fact, in the
limit $m\to 0$ we have
\be
\partial_t z = -K\partial^4_x z -\nu\partial^2_x z
\label{eq-lin}
\ee
whose solution is $z(x,t)=\exp(\omega_q t)\cos(qx)$ with
$\omega_q =\nu q^2 -Kq^4$. An up-hill current means that $\jes$ has the same 
sign as the slope, so $\nu$ is positive and the flat surface is unstable
($\omega_q >0$) against modulations of wavevector 
smaller than $\bar q=\sqrt{\nu/K}$; the
instability appears after a typical time of order 
$t^*\simeq (\nu {\bar q}^2)^{-1} = K/\nu^2$.

The later evolution of the surface depends on the nonlinear form of the
unstable current $\jes(m)$. By taking the spatial derivative of 
Eq.~(\ref{eq-z}), we obtain
\be
\partial_t m = \partial^2_x (-j) + \partial_x (\delta F)
\ee
and a parallel with a phase ordering process is easily done, once we remark
that the current can be obtained by a pseudo free energy ${\cal F}$:
\be
j =-{\delta {\cal F}\over\delta m}~,~~~~{\cal F}[m] = \int\dx\left[
{K\over 2} (\partial_x m)^2 + V(m)\right]~,~~~~V'(m) = -\jes(m)~.
\ee

The instability of the flat surface $(\jes'(0)>0$) means that the potential
$V(m)$ has a maximum in $m=0$ ($V''(0)<0$). 
Contiguous regions of increasing and opposite
slope are formed. 
The usual phase ordering process is obtained when $V(m)$ has 
the classical double well form: $V(m)=-(\nu/2)m^2+(\nu/4m_0^2)m^4$,
corresponding to a current $\jes=\nu m(1-m^2/m_0^2)$.
After the slope has attained a fraction of $m_0$ the 
dynamics enters in the nonlinear regime: the wavelength $L$ of the profile
increases in time (coarsening process) and the slope saturates 
to the constant values $\pm m_0$.
The coarsening law is known to be logarithmic~\cite{Langer} ($L(t)\sim\ln t$) 
in absence of shot noise and a power law~\cite{KOKM} 
($L(t)\sim t^{1/3}$), in presence of it.

The aim of the present paper is to analyze the {\em deterministic} 
$(\delta F(x,t)\equiv 0)$ growth process when
$V(m)$ has no minima, corresponding to the absence of zeros at finite slopes
in the current $\jes$. 
We will consider the class of currents defined by
\be
\jes = {\nu m\over (1+\ell^2 m^2)^\alpha} \hbox{~~~~~~~with~~~~}\alpha \ge 1
\label{1class}
\ee
and the corresponding models will be termed $\alpha$-models.

Model-1 has been studied numerically by Hunt \etal \cite{Sander} 
and they found a coarsening
exponent $n\approx 0.22$ ($L(t)\sim t^n$). $\alpha$-Models without noise
have been
studied analytically by Golubovi\'c~\cite{Golub} through scaling
arguments and he finds $n=1/4$
irrespectively of $\alpha$. Finally, qualitative considerations based on
noise effects~\cite{Privman-Tang} give $n=1/(2/\alpha+3)$, 
i.e.  $n=1/5$ for model-1.

Our analytical approach is based on the linear stability analysis of
the stationary configurations $j[m(x)]\equiv 0$. This way, the 
finding of the coarsening exponent $n$ passes through the determination
of the lowest eigenvalue of the operator $(-\partial^2_x)\hat H$, where
$\hat H$ is the Hamiltonian corresponding 
to a particle in a periodic potential~\cite{Langer}. 

Before proceeding we render adimensional the growth equation by 
rescaling $x$ with $1/\bar q$, $t$ with $t^*$ and $z$ with $1/\bar q\ell$:
\be
\partial_t z = -\partial_x j~, 
~~~~~j=m'' + {m\over (1+m^2)^\alpha}~.
\label{z_eq}
\ee

Stationary configurations are the solutions of the differential equation
$j[m(x)]=m'' + \jes(m)\equiv 0$. 
Therefore they correspond to the periodic orbits
of a particle in the potential 
$-V(m)=-[1/2(\alpha -1)](1+m^2)^{1-\alpha}$ 
for $\alpha>1$ and in the potential $-V(m)=(1/2)\ln(1+m^2)$ for $\alpha=1$.
In the former case the potential is upper bounded and the solution
corresponding to the boundary conditions $m\to\pm\infty$ when $x\to\pm\infty$
does exist, while it does {\it not} for $\alpha=1$ because the corresponding
energy would be infinite. Stationary solutions may be labelled with their
period, i.e. the wavelength $L$: $m_L(x)$. 

Let us now perform a linear stability analysis around these stationary and 
periodic solutions: $m(x,t)=m_L(x) + \psi(x,t)$. It is easily found that
\be
\partial_t \psi = \partial_x^2\left[ -\psi''(x,t) + U_L (x)\psi\right]~,
\ee
where $U_L (x) \equiv -\jes'(m_L(x))$. By putting $\psi(x,t)=\phi(x)
\exp(-\epsilon t)$ we obtain
\be
(-\partial_x^2)\left[ -\phi''(x) + U_L (x)\phi\right]
\equiv D_x\hat H\phi (x) =\epsilon\phi~.
\ee

Negative eigenvalues mean that $m_L(x)$ is linearly unstable and
this induces the coarsening process; moreover, 
$\epsilon(L)\to 0^-$ when $L\to\infty$. The dependence
of the ground state (GS) energy on the distance $L$ determines 
the time scale of the coarsening process: $t\sim 1/|\epsilon(L)|$.
For the moment we will assume $D_x\equiv 1$, i.e. we will consider the
{\em nonconserved} model: $\partial_t m = -\delta {\cal F}/\delta m$.
 
First of all we observe that in the limit of large $L$ the energy shift
$\epsilon(L)$ for the periodic potential is equal (up to a
numerical factor) to the shift for a single couple of potential
wells~\cite{pp2b}.
The solution of the problem is given~\cite{ll} in terms of
$\phi_0$ and $\phi_1$, respectively the ground state for the single 
well $U_1(x)$,
centered in $x=L$, and for the double well $U_2(x)$, centered in $x=\pm L$.
In fact the Schroedinger equations are:
\be
-\phi_0'' + U_1\phi_0 = 0~~(a)~~~~~~~~
-\phi_1'' + U_2\phi_1 = \epsilon\phi_1~~(b)
\label{eq-sch}
\ee
and by evaluating the quantity 
$\int_0^\infty\dx[\phi_1\times(\ref{eq-sch}a)-\phi_0\times(\ref{eq-sch}b)] =0$,
we obtain 
\be
\phi_1(0)\phi_0'(0) = - \epsilon\int_0^\infty\dx\phi_0(x)\phi_1(x)~,
\label{landau}
\ee
where we have made use of $U_1=U_2$ for $x>0$.

Before proceeding we must determine the asymptotic expressions of $\phi_0(x)$ 
and $\phi_1(x)$. The potential $U(x)=-\jes'(m)$ is given,
for $\alpha$-models, by
\be
U(x) = {(2\alpha-1)m^2 -1\over(1+m^2)^{\alpha+1}}
\to {(2\alpha-1)\over m^{2\alpha}}~.
\label{UU}
\ee

The asymptotic behaviour of the single-mound profile is obtained by
integrating the equation $m''(x) +\jes(m)=0$ and taking the limit
$x\to\infty$\, :
\be
(1/2)(m')^2 - V(m)=0 ~~~\Rightarrow ~~~{dm\over dx}\approx 
{1\over\sqrt{\alpha -1}} {1\over|m|^{\alpha-1}}~.
\ee
The result $m^\alpha(x)\approx (\alpha/\sqrt{\alpha-1})x$, when inserted in 
(\ref{UU}) gives:
\be
U(x)\approx {(2\alpha-1)(\alpha-1)\over\alpha^2} {1\over x^2}\equiv
{a\over x^2}~,
\ee
with $a$ increasing between $a=0$ (for $\alpha=1$) and $a=2$ 
(for $\alpha=\infty$).

The solution of the Schroedinger 
equation~(\ref{eq-sch}a) for $U_1(x)\approx a/(x-L)^2$ gives a
{\it power-law} decaying wavefunction ($\phi_0(x)\sim |x-L|^{-\beta}$), with
an exponent $\beta = (1-1/\alpha)$.

If $\alpha\le 2$ then $\beta\le 1/2$ 
and therefore the GS $\phi_0(x)$ of the single well is not a 
bound state, since $\int_{-\infty}^{\infty}\dx\phi_0^2(x)=\infty$.
On the other hand, for 
$\alpha>2$ $\phi_0(x)$ is a bound state and $\phi_1(x)$ can be
approximated~\cite{ll} 
with the expression $\phi_1(x)=[\phi_0(x)+\phi_0(-x)]/\sqrt{2}$.
This way, from~(\ref{landau}) we easily obtain the relation
\be
\epsilon\simeq -2\phi_0(0)\phi_0'(0) \approx - L^{-(2\beta+1)} ~~~~
[\alpha>2 \hbox{~and~} D_x=1]~.
\label{dalandau}
\ee

If $\alpha <2$, we can put 
$\phi_1(x)=[\tilde\phi_0(x)+\tilde\phi_0(-x)]/\sqrt{2}$ where $\tilde\phi_0$
is a generalization of $\phi_0$ to a negative eigenvalue:
$-\tilde\phi_0''(x) + (a/x^2)\tilde\phi_0(x) = \epsilon\tilde\phi_0(x)$.
In fact, even if $\phi_0$ is not a bound state, $\phi_1$ is bounded, 
because the
GS energy $\epsilon$ is strictly lower than $U_2(\pm\infty)=0$.
The previous expression for $\phi_1$ may be used even if $\phi_0$
itself is bounded (i.e. for $\alpha>2$) and the result for the
coarsening exponent does not change.

The asymptotic expression for $\tilde\phi_0$ is $\tilde\phi_0(x)=\sqrt{x}
K_\mu(\sqrt{|\epsilon|}x)$
where $K_\mu$ is the modified Bessel function of order $\mu=\beta-(1/2)$.
The function $\tilde\phi_0$ decays as a power-law ($\tilde\phi_0(x)\approx
|\epsilon|^{-\beta/2-1/4} x^{-\beta}$) if $a/x^2\gg|\epsilon|$ and
exponentially ($\tilde\phi_0(x)\approx|\epsilon|^{-1/4}
\exp(-\sqrt{|\epsilon|}x)$)
in the opposite limit, $a/x^2\ll|\epsilon|$.
Eq.~(\ref{landau}) now writes
\be
\epsilon\int_0^\infty\dx\phi_0(x)\tilde\phi_0(x) =
-2\tilde\phi_0(0)\phi_0'(0)~~~~[\alpha\le 2 \hbox{~and~} D_x=1]~,
\label{landau2}
\ee
where $\tilde\phi_0(x)$ depends on $\epsilon$. Let us remark that the
integral $I$ on the left hand side does converge even if $\phi_0$ is not a
bound state.

The evaluation of the two sides of Eq.~(\ref{landau2}) is a bit lengthy
and we report here the result only:
$|\epsilon|\ln(1/|\epsilon|)\sim 1/L^2$ if $\alpha =2$ and 
$|\epsilon|\sim 1/L^2$ if $1<\alpha <2$.
In Fig.~\ref{fig1} we compare the analytical results for the 
exponent characterizing the energy shift
$|\epsilon(L)|\sim L^{-1/n}$ with those obtained 
through its direct numerical evaluation~\cite{note2b} and the agreement
is very good.

Therefore, for the nonconserved model we can conlude that:
\be
{\tt [nonconserved]}~~~~
n={1\over 2}~~~(1<\alpha\le 2)\hbox{~~~and~~~}n={1\over 3-2/\alpha}~~~
(\alpha>2),
\label{nonconserved}
\ee 
with a logarithmic correction for $\alpha=2$ ($L\sim (t/\ln t)^{1/2}$).

The reason why the coarsening exponent $n$ keeps constant for $\alpha<2$ is
the following: if $\alpha>2$ the single-well wavefunction is a bound state,
the integral $I$ is a constant while the `superposition' between
$\phi_0(x)$ and $\phi_0(-x)$ (that is to say the right-hand side of 
(\ref{landau2})) decreases at increasing $\alpha$, which implies a decreasing
$n$. Conversely, when $\alpha<2$ the integral $I$ becomes 
$\alpha-$dependent and decreases with $\alpha$: these
dependence counterbalances the reduction of the right-hand side of 
(\ref{landau2}). 

For the {\em conserved} growth model, $D_x=-\partial_x^2$
and Eq.~(\ref{landau}) must be replaced by a more complicated expression.
It has not been possible to carry out a rigorous calculation because
$[\phi_1 \times D_x \hat H\phi_0 - \phi_0 \times D_x \hat H\phi_1]$ is no more
integrable. Nonetheless, there are strong indications
that the right-hand sides of (\ref{dalandau},\ref{landau2})
acquire a factor $L^{-2}$: the origin of this scaling factor is that 
$\phi_0(x)$ has a power-like behaviour (and therefore derivation
corresponds to divide by $x$) and also that $U(x)\sim x^{-2}$.
Furthermore, since we need the single well wavefunction, corresponding to 
a zero energy, a solution of the Schroedinger equation $\hat H\phi(x)=0$
is also solution of $D_x\hat H\phi(x)=0$.

As a consequence of such factor, the coarsening exponent for the conserved case
is easily obtained from the nonconserved one: $(1/n) \to [(1/n) +2]$.
Therefore:
\be
{\tt [conserved]}~~~~
n={1\over 4}~~~(1<\alpha\le 2)\hbox{~~~and~~~}n={1\over 5-2/\alpha}~~~
(\alpha>2)~.
\label{truth}
\ee

In order to check numerically the validity of the results
reported in Eq.(\ref{truth}) and therefore the
dependence of the coarsening exponent
$n$ on the parameter $\alpha$, detailed numerical
simulations have been performed. In particular,
we have numerically integrated  equation~(\ref{z_eq}) 
by employing a pseudo-spectral time splitting code~\cite{algo}. 

The values of $L(t)$, whose log-log plot gives the exponent $n$,
are evaluated through the power spectrum (PS) of $z(x,t)$: 
the weighted average of the wavevectors corresponding to the most
relevant components of the PS is $2\pi/L(t)$.
A different method
using the spatial correlation function gives consistent results.
In Fig.~\ref{fig2}, the numerical findings for $n(\alpha)$ by direct
integration of Eq.~(\ref{z_eq})
are shown together with the theoretical expression 
(\ref{truth}) and a good agreement is found.

In conclusion we have found the analytic expression for the
coarsening exponents $n(\alpha)$, both for the nonconserved
model, Eq.~(\ref{nonconserved}) and for the conserved one (growth model),
Eq.~(\ref{truth}).
Coarsening varies with $\alpha$ and it is not logarithmic (i.e.
$n =0$) even for $\alpha=\infty$.

\ack
We have benefitted a lot by useful discussions with several colleagues:
A. Crisanti, C. Godreche, J. Krug, M. Moraldi, W. Strunz and S. Taddei. 
PP gratefully acknowledges financial support from Alexander von Humboldt 
Fondation.

\Figures

\begin{figure}
\caption{Analytical (full line) and numerical (crosses) values for the
exponent $1/n$ governing the asymptotic energy shift $|\epsilon_2|\sim
1/L^{1/n}$ (nonconserved model).}
\label{fig1}
\end{figure}

\begin{figure}
\caption{Coarsening exponent $n$ for the conserved model. 
In the inset we enlarge the small $\alpha$ region.
Full line is the
analytical result (Eq.~\protect\ref{truth}). Points are the exponents found
integrating numerically Eq.~\protect\ref{z_eq} for  
a system size $M=1024$ (spatial resolution $\Delta x = 0.25$)
and a total time $ 400,000 < T < 1,600,000$
(time step $\Delta t = 0.05$). A few tests have also been done with a 
smaller time step ($\Delta t = 0.025$) and longer chains ($M=2048$-$4096$),
obtaining consistent results.
Bars indicate the numerical fit errors.
}
\label{fig2}
\end{figure}


\begin{thebibliography}{11}
\bibitem{pp2b}
If $\epsilon_2$ is the energy shift for a couple of wells, the shift
$\epsilon_n$ for $n$ wells is $\epsilon_n = 2\epsilon_2 (1-1/n)$.
\bibitem{note2b}
We have considered the single well potential $U(x)=-E_0$ for $|x|<1$
and $U(x)=a/x^2$ for $|x|>1$, where $E_0(a)$ is chosen so as to provide
a zero GS energy. Afterwards the energy $\epsilon_2$ of the double well 
problem may be determined by joining the solutions of
the Schroedinger equation in each separate ``piece" of the potential
$U_2(x)$.
\end{thebibliography}
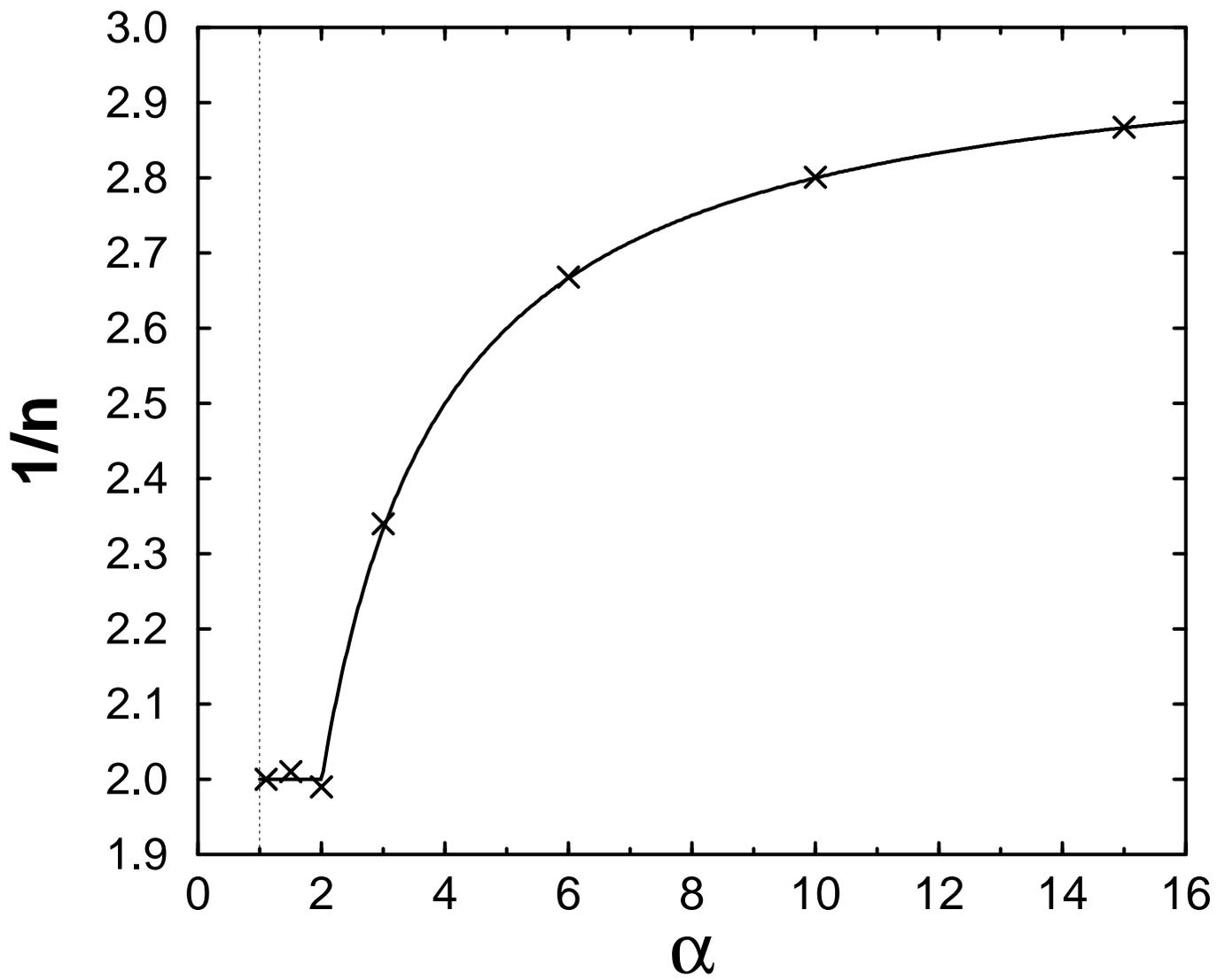
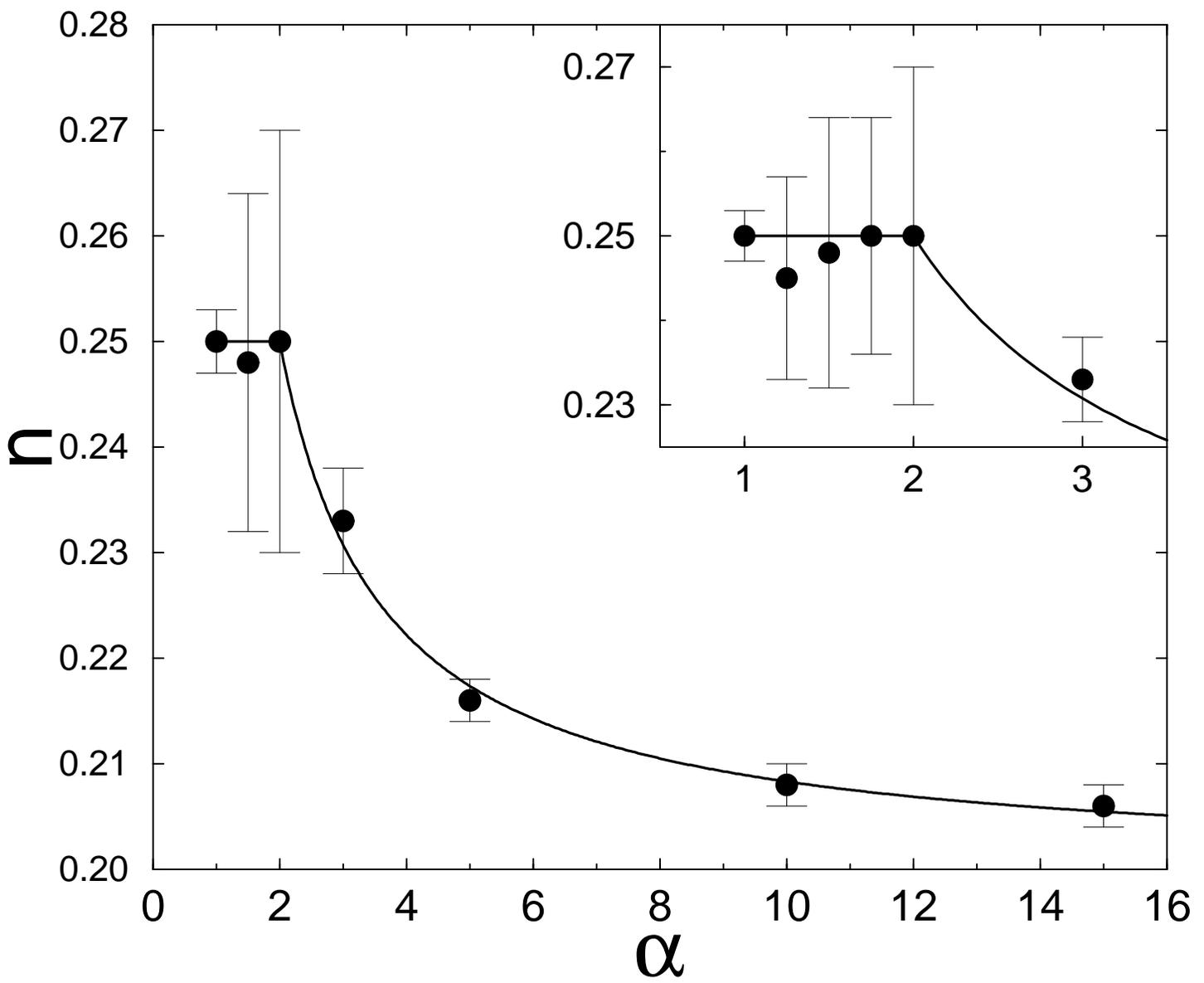
\end{document}